# Because we care: Privacy Dashboard on FirefoxOS

Marta Piekarska
Security in Telecommunications
Technische Universität Berlin
and Telekom Innovation Labs
Email: marta@sec.t-labs.tu-berlin.de

Yun Zhou
Assessment of IP-based Applications
Technische Universität Berlin
and Telekom Innovation Labs
Email: Yun.Zhou@telekom.de

Dominik Strohmeier
Mozilla Corporation
Email: dstrohmeier@mozilla.com

Alexander Raake
Assessment of IP-based Applications
Technische Universität Berlin
Email: Alexander.Raake@telekom.de

*Abstract*—In this paper we present the Privacy Dashboard – a tool designed to inform and empower the people using mobile devices, by introducing features such as Remote Privacy Protection, Backup, Adjustable Location Accuracy, Permission Control and Secondary-User Mode. We have implemented our solution on FirefoxOS and conducted user studies to verify the usefulness and usability of our tool. The paper starts with a discussion of different aspects of mobile privacy, how users perceive it and how much they are willing to give up for better usability. Then we describe the tool in detail, presenting what incentives drove us to certain design decisions. During our studies we tried to understand how users interact with the system and what are their priorities. We have verified our hypothesis, and the impact of the educational aspects on the decisions about the privacy settings. We show that by taking a user-centric development of privacy extensions we can reduce the gap between protection and usability.

## I. INTRODUCTION

User activities are shifting from personal computers to smartphones. Studies show that more Internet accesses are made from mobile devices than from desktops, especially in emerging markets [1]. It is often said, however, that users do not care about their privacy. On the other hand, when not in the virtual world, they tend to preserve their personal information – hide their diaries, avoid disclosing certain information in front of strangers, protect their correspondence. Our hypothesis is that when it comes to the mobile reality, the problem lays in the lack of intuitive privacy protection mechanisms and education of the users. This hypothesis seems to be supported by the public reaction to the PRISM [2] and articles on surveillance published in mass media, like "Der Spiegel" [3] or "The Guardian" [4].

Existing tools do not give full and easily understandable information about good privacy practices, nor do they allow to control the phone sufficiently [5]. Our aim is to help users by increasing the perception and awareness of potential risks, along with designing mechanisms to protect against information leakage. In this paper we will discuss an approach to design and development of privacy-preserving tools. We call it the "User-centric privacy development", which is further discussed in the next section. In Section III we discuss work in both academia and the industry on the topic of mobile privacy. Then, in Section IV we will move to the description of the study we have performed to find the needs and expectations of users with respect to the privacy preservation on mobile phones. In Section V we will present the design that we have developed to meet the results of the initial study. The Privacy Dashboard, as we call it, was implemented to work with Firefox OS, and its details will be shown in Section VI. Next, in Section VII we describe the verification study. Finally in Section VIII we will talk about the results and insights we have gained from our work – user behavior and patterns when using tools like ours. We finish in Section IX with some conclusions.

The contributions of the paper, are as follows: (1) We formulate and realize a user-driven, approach to the design of multi-feature privacy tool.(2) We identify what is the gap between the users' expectations and the actual data protection on the system. (3) To this end we design and implement features that address the problems (like Remote Privacy Protection, Backup, Adjustable Location Accuracy, Permission Control and Guest Mode) that we combine in the Privacy Dashboard, which we then (4) evaluate with extensive user-studies.

## II. USER-CENTRIC APPROACH

User-centered design (UCD) emerged from Human-Computer Interaction (HCI) and is a software design methodology for developers and designers. It helps them make applications that meet the needs of their users [6]. The approach of UCD is created in three main iterative steps as requirement for exploration, design research and evaluation.

We found that people have a certain expectation of privacy, but they don't know how to enforce it on their smartphones. We adopted UCD to bridge the gap between people's expectations and the actual workings of the device. In order to achieve this goal, we make iterations of user studies and technical work as shown in Figure 1. We have used the approach to create a more generic Privacy Dashboard. The same scheme, however, can be used to develop single features. We have started off by initial user study where we gather the guidelines for the planned solution. We check what do they expect from the tools they get; how do they use the device; how does their understanding of the system differ from the reality; what is their attitude to the problem. Based on the initial studies, we can create a set of paradigms that the system should, but is not, fulfilling. Next, rather than focusing on the technical details, we want to provide multiple versions and possibilities that address the same problems, and verify how people feel about them. Having an extensive prototype implementation and previous results, we can create the questionnaire and design the test scenarios, to give us in-depth review of our approach. As a result of the second phase of user studies we improve the model and finalize the tool.

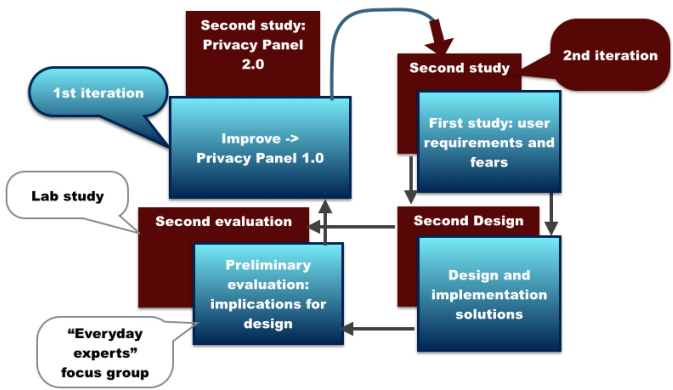

Fig. 1: UCD development of Privacy Dashboard

## III. RELATED WORK

We would like to present work done in the academia in the fields of both user studies and the tools, as well as a brief analysis of apps available on the market for Android and iOS.

### A. On users perception

To build with a user-centric approach we use the knowledge gathered in the field of information and data sharing, as well as people's awareness of privacy issues. In [7] the authors present insights into up-to-date mobile device sharing behavior. Their work analyzes users' concerns with respect to data and device sharing. This gives basis for creating our own user studies for Privacy Dashboard, and Guest Mode in particular.

Another work was presented by Keng et al in [8]. They have demonstrated the feasibility and benefits of identifying the causes of leaks from a user's point of view, which they call *mobile forensics of privacy leaks*. Not only can they correlate user actions to leaks, but also report the causes from a user-oriented perspective.

Insights into users' behavior and perception of mobile privacy are given in [9], [10], [11], [12]. We have based our plans and approach on these works.

A study presented in [13] by Keith et al. proposes and tests a more realistic experimental methodology designed to replicate real perceptions of privacy risk and capture the effects of actual information disclosure decisions. Another paper by Jialiu Lin, [12], addresses this gap through the study of mobile app privacy preferences with the dual objective of both simplifying and enhancing mobile app privacy decision interfaces.

### B. On tools

Academia is addressing the issue of mobile privacy. There are, however, very few tools available which would allow to control the privacy settings of the phone, or extend the level to which we can influence the data exchange.

In [14] the authors describe a "Privacy Panel App" in which they represent computed ratings of privacy impact. These are offered in a graphical user-friendly format and help the users in defining policies based on them. This is an interesting idea of how to create a tool that would be useful for an average user, as well as allow him to get an overview of what does his phone do. However, the authors do not provide methods to empower the user, instead showing him what is happening on his phone.

Works like [15], [16], [17], [18] present solutions to partial problems – location blurring, secure communication, encryption on mobile devices. However these are scattered around and would be hard to combine into a single tool.

### C. System Settings on Android and iOS

With the introduction of iOS 8, Apple decided to take a big step and change their privacy policy. In an official statement they have promised to no longer unlock iOS 8 devices for law enforcement [19]. This is a change in the practices of a company that was well known to collaborate with all sort of government institutions [20]. They have also been fixing a set of vulnerabilities that have been detected by external experts [21]. There is an extensive privacy tab, where users can choose which apps have access to which sensitive resources – Location Services, Contacts, Photos, Microphone, etc. There is also a possibility of limiting the Ad Tracking and activating a Find my Device feature. User can choose what, if anything, will be uploaded to the iCloud – the backup mechanism for iPhone. However, the control given to the user is very binary – either you reveal location to an app or you block it – there is no method of "fuzzing" the position of the phone. Also there are services like "Share my location" or "Find my Friends" and it is quite hard to find the place where one could disable the feature. Moreover, there is no explanation as to what each element does, and no information on why an app would need access to certain information, which may actually fully justify the request.

Android is not a particularly privacy friendly system. The privacy settings built into the OS are very limited and well hidden. There was a moment, in version 4.3 of the system, when an app called App Ops [22] was available. It provided information about what the phone did and what resources did it access. With Android permission model, however, there is no way to cherry-pick the elements app can access. The decision is made during the installation process and is a binary one. Notably, App Ops did not provide any ability to control whether an application should have access to the Internet. It was also not possible to prevent apps from uniquely identifying a device, or its owner, via third party accounts. This means that an app could still tie together all of the account identities on the device and access IMEI and other unique device identifiers with the appropriate permissions.

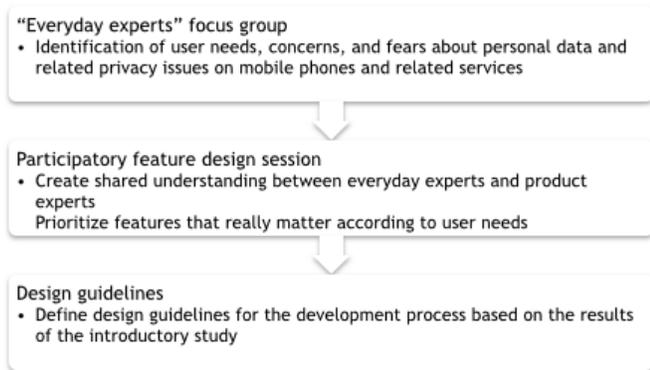

Fig. 2: The concept of the introductory study and the key results of each step for the user-centered design process

## IV. INTRODUCTORY STUDY

We have presented the literature overview and the general concept of user-centric privacy design in previous sections. Let us now move to the discussion of the realization of such model.

### A. Methodology

The introductory study was split into two parts and a result summary, as presented in Figure 2. We recruited twenty-four users over social media channels to join a focus group. We chose the participants as "every day experts" – people between 20 and 40 years old, that were evaluated to have average knowledge about mobile phones and data privacy. The group was selected to eliminate people with highly specialized technical background. The goal of this first session was to identify needs and fears, motivation barriers as well as general comprehension and mental models that users have about privacy on mobile phones. All "everyday experts" owned a smartphone for at least 1 year and used a variety of services like email, instant messaging, or navigation regularly. The first session was conducted some days prior to the second phase to allow for analysis of the group work as input to the next step.

In the second part of the study, everyday experts teamed up with product and engineering experts from the field of security and privacy. First, we wanted experts to hear from users about privacy and learn how users' mental models work to be able to use this input for the user-centered development process. Second, an everyday expert and one member of the Product group paired up to develop communication about privacy features.

### B. Results

Users reported having dilemmas with respect to sharing their private data. However, they tend to accept non-transparent privacy policies rather than taking actions or opting-out from a service. Three principals describe the main needs that users have about privacy: feeling in control, ease of use, and taking actions.

From a user perspective, sharing data on the Internet is an ambivalent issue, as we have seen in our study. On one hand, users believe that a service will only work, or at least work better, if they disclose their personal data to it or accept the private data collection. On the other, people are aware of the threats that sharing data poses, thus they tend to be uneasy about it. We could identify three main categories of complaints that users had:

1) Users commonly reported that they do not trust apps that do not state clearly why they request access to the data, or what the consequences of such sharing are ("I am worried when I need to share data, that it absolutely irrelevant to a service that I want to use.", "I feel uncomfortable when services use my data for creating personalized advertisement.", "Why does a service need to check my email contacts at first time use?").

2) Fears are caused by a brand that offer the service and its image ("I am worried when brands whose services I use get negative press in the media.", "[This company] is known for bad habits against their employees, I assume they also take user data best practices not too serious.").

3) The third category, however, comes from lack of education. People worry that they do not know and understand how an app works or how to use it.

While working with both regular users and the technical experts, we were surprised to see that the motivation not only to question the validity of the data sharing, but also to search for alternative services is low: "Eventually, there is no chance not to use this service, so I disclose all the data knowing that this might be wrong". What we have seen is that users start to accept that the Internet does not work without sharing personal data ("Nowadays in the Internet you cannot expect to know who is on the other side of the screen. That's part of using the Internet."). Moreover, people do not value their personal data ("I don't care if somebody gets all my data from this service.", "I have nothing to hide."). Finally, most participants confirmed, that they prefer to outsource to the technical experts the problem of data privacy. They are willing to trust the "experts" to create systems that protect their information. They are willing to give up part of the usability for privacy, but not so much that they would not be able to use a service any more.

## V. DESIGN

We identified problems that were common to certain groups – basic users, technology aware people and advanced. Depending on the attitude and knowledge, the expectations of the system were different – some people would prefer to sacrifice privacy in order to be easily reachable, while others would much rather limit the information sharing to the minimum.

*A. Problem statement*

In order to verify our user-centric approach we decided to focus on the *basic user*. People in this group have little to no technical knowledge, might be first-time smartphone users, believe that either technology is scary or very safe and protective. The common threats we have initially identified are two-fold: on one hand people feel like they do not have enough knowledge about good privacy practices, on the other – even if they do, they do not feel in control over the phone. Our report showed that:

- Users want personal information to be safe and secure, especially when a device is lost or stolen.
- Users want to know their data will not be inadvertently exposed to unauthorized people.
- Theft of devices, resulting in loss of personal data, is a common problem on emerging markets.
- Users want control, but also need help in setting up and managing their devices and data.

*B. Requirements*

In a user-centric solution we need to assure high technical quality and be aware of the risks. However, it is also very important not to approach the problem from strict security research point of view sacrificing the usability. There are elements that have to be accounted for when designing our tool:

- We will be working with low-in-computational power devices
- The tool has to be as protective as possible, within the legal limits in a variety of countries.
- As a result – the tool will not protect from legal interception, or other governmental surveillance methods.
- The main purpose is not only to empower the users, but also to make them aware of the threats – so that they can adjust their everyday practices.
- The tool has to ensure a high quality of technical solutions – half-baked privacy is worse then no privacy.

*C. Capabilities*

To address the issues described in Sections V-A and V-B we decided that the tool should offer the following capabilities:

*1) Guided Tour:* One of the first things that showed up in our research was a lack of understanding and knowledge on good privacy practices. Almost all people, despite their initial technical skills, feel lost. The Guided Tour (GT) is a set of explanatory screens that discusses the capabilities and offers explanation on what features Dashboard introduces. After completing the Tour, user can enter the main panel and proceed with the set-up. Later, at any moment if one feels like he needs help, he can re-enter the Guided Tour on chosen topic.

*2) Lost or stolen device control:* First and foremost the user should be able to track his device – when one loses the phone out of sight, he should command it to report back where it is. Second, until the device is found, it should lock itself remotely with the passphrase – so as long as it is out of user's control, no one else should be able to access it. Also despite the locating abilities, it has to be possible to wipe the data from the phone – in case the phone gets permanently stolen, the best we can do is protect our privacy.

*3) Geolocation privacy:* The Directive on privacy and electronic communications [23] states that the location-based services (LBS) must be permission-based. Thus, since 2002, the user must opt-in in order to use the LBS. A common solution is to turn it off. But that makes some apps, navigation for instance, unusable. Moreover, in many cases, the application highly benefits from roughly knowing where the user is, but doesn't necessarily need to know his precise coordinates.

*4) Controlling apps:* Users have no control what information is shared and when. Even if they take part in the decision process – while installing an Android App, user needs to agree to the permissions it requests – they never know when the actual accesses are happening and why. A study by Consolvo et al [24] around location sharing, showed that the decision people make about the permission granting depends on the context: who was requesting it, why did they ask for it, and how detailed the information should be. The Privacy Dashboard should thus introduce a good overview of what each app is doing and when, to provide basis for an informed decision on allowing or withdrawing access to certain resources.

*5) Sharing devices with friends and family:* It is not very rare that one wants or needs to share their phone with someone else. Nevertheless, the data we store on mobile devices should not be exposed to anyone random. Although the concept of a "Guest Mode" is well established on PCs, on mobile phones we lack some kind of a Secondary-user Mode, which would allow the owner of the device to borrow it to others, without worrying about his personal data. Limited functionality is offered by Apple [25] (under the name "Guided Access") and Android [26].

## VI. IMPLEMENTATION

It is not the scope of this paper to present each solution in great detail – every one of them carries some sophisticated challenges. As we focus on the general approach of user-centric tool development, this section will include the implementation details that are important to the approach. We have decided to work on Firefox OS, as it is an open source system, available on phones in the emerging markets. It allowed us to adjust and change the system to our will. Moreover we wanted to test our solution with the people who have less experience with smartphones and the privacy threats they pose.

*A. Guided Tour*

The Guided Tour was mostly a communication challenge. We had to investigate how do we explain the technical aspects of the Privacy Dashboard to the average user. What language should be used in order to inform, and not to scare. We have done extensive literature study on the decision making process and the psychology of judgment. Work by Mellers et al. [27] provided great guidance in the field.

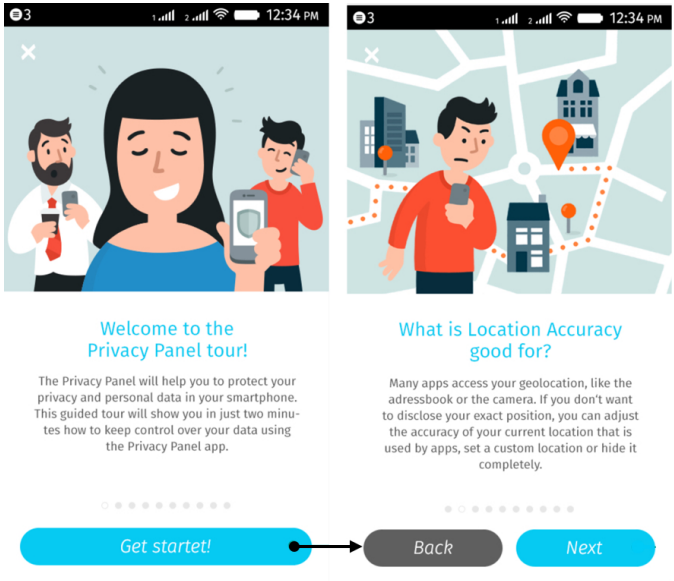

Fig. 3: First couple wireframes of Guided Tour. Each panel consists of a graphical illustration that provides suggestion about the topic, a title and a short explanation. Instead of simply describing the feature, the Guided Tour gives the consequences of the decision.

There are three main elements involved in the decision making: the risk perception, the risk attitude and the emotions. It is an obvious, yet important realization: not only emotions have high impact on decisions, but also the results of our decisions have powerful effect on our emotions. Surprisingly framing effects, stimulus contexts, environments, and response modes can have a big influence on the decision making. Of course, rules and habits also play a great role in the process.

All this gave us a good understanding on how to educate people about good privacy practices. It was clear that the message has to be short and simple. It has to carry some emotional load, and give the sense of participating in something interesting and exciting. Additionally suggestions that guide people should be made, giving them a reference point.

We decided to introduce characters that will be the guides during the exploration of privacy settings. Each screen of the Guided Tour consists of a graphic that shows one of our characters in a situation that gives an intuition about the explanation. The text is short and serves the purpose of discussing the feature, giving reasons why it is important and consequences of the decision.

We introduced two versions of the Guided Tour: one that was interactive and allowed for adjusting the settings while reading about them, and second where the user could only read the explanation. During the second study the users chose the latter, as can be seen in the Figure 3. We also included the possibility of sharing the privacy settings with other members. This allows users to follow the guidance from the people they trust – be it their IT friend or recommendation of the favorite social network.

### B. Remote Privacy Protection

When we ask people what are their anxieties in mobile world, many mention losing their phone. However, when we talk about privacy in the context of existing tracking solutions, users have doubts about registering in order to connect your device to the online identity: both iOS and Android require the owner to create an account during the first time use. During our initial study, we found one statement very inspiring: "Trust is something that is earned, not enforced. Why do everyone take my trust in technology for granted". Coming from the account-less approach we have created a tool that allows manipulating the device with the use of short text messages. When initially starting the Remote Privacy Protection (RPP) feature, user is asked to create a passphrase. This code is later used as the protection mechanism to block changing of the settings, and to identify the owner of the device. Next user will enter simple view where he can choose which commands does he want to activate through the SMSes. The possibilities are: locate, lock, ring and wipe. When the phone is lost, user can use any device he has at hand, including some free online services, and send a message to his number. The message should be formatted as:

^rpp \s(lock—ring—locate) \s([a-z0-9]1,100)$/i

Every time there is an incoming message, an event listener in the Remote Privacy Protection checks if it starts with the keyword and if it does – it will handle the verification process. When the command is correct, and the passphrase matches the one set by the user, device will act accordingly:

- Lock – phone turns on the lock screen with a passphrase.
- Locate – the phone sends out the current geoposition and locks itself. It is still responsive to new locate commands.
- Ring – the phone locks itself and activates the ringer with the highest possible volume.

As any approach this method has its drawbacks, most of which come from the fact that we do not want to limit the communication only to phones with the app installed, but allow for sending out the SMS from virtually any device. First problem is the visibility of the incoming text messages. Although most devices support hidden SMSes, none of them does it by default. That is why we cannot rely on this feature. It means, however, that the incoming text will be displayed on the stolen phone, hinting the attacker that someone is manipulating it. It also means that we cannot set passphrases in the commands, as the person holding the phone would be able to learn it. This, and the problem of man-in-the-middle attack, could be solved by encrypting the message. Then the attacker would only see random characters. Again, however, if we want to use standard Text App, independent from the OS, there is no way of introducing an encryption mechanism on the sender side. Last method that comes to mind is usage of an authenticatior that generates one-time passwords. One example could be the Google Authenticator[1]. However this again creates the problem of synchronization and system dependance. In order to create a one-time password, user would need to have a phone with the authenticator app installed. We are in a process of creating a solution to that problem, however it is too early to discuss the possible mechanisms.

---
[1]https://play.google.com/store/apps/details?id=com.google.android.apps.authenticator2

We handle these problems in a simple way. The passphrase has to be set up during the first time use, and can be as long as 100 characters. Once the device is found and the correct passphrase is entered, the app resets itself and asks for a new passphrase. The new entry is compared with the old one so that the user cannot enter the same value twice in a row.

*C. Backup*

Giving the users a choice, whom to trust with data, led to another improvement in privacy protection. We have created a backup mechanism that allows to choose where the data would be stored. With the Privacy Dashboard, user is able to pick the location – be it default Mozilla server, some popular storage provider, or his own computer. This solution has little technical challenges, and rather extends the functionality.

*D. Adjustable Location Accuracy*

The concept of blurring the position of a user, so that not every application has his precise coordinates is very appealing to most people [28]. We have to work within the legal limits, so we cannot influence the way the phone reports its position to the network operator. We can, however, change the API in such a way that the location is adjusted according to the settings, and we also can try using different proxy servers to hide the IP address.

In the Privacy Dashboard, the user can choose the accuracy of the location services on per-application basis:

- **Turn Location Off** allows the user to not give any location data at all.
- **Give Precise Location** leaves the system without any changes.
- **Choose a Position** allows the user to fix his position to a set of coordinates. We provide a list of predefined values and a search that allows to find a City or Country (where the coordinates are set to the center of mass of the place). Additionally the user can enter custom latitude and longitude.
- **Blur by X km** here the user chooses the distance by which his position will be randomized. The choice is flexible and can vary from 1 to 500 km.

To avoid some side channel timing attacks in every case we allow the geolocation API to obtain the position. Next, we verify what app is calling the API and check what is the setting for it, and adjust the response the API would give. As the application is not aware of the process and gets valid data, app cannot find out if the blurring or fake location is turned on. The basic concept is presented in the Figure 4.

In order to introduce the blurring we divide the world into a grid. Its size is defined by the user, and can be different for every app. We then determine the cell in which the phone is, and return the middle of the cell. As long as the user stays within that cell, the position reported to the app will not change. When he moves to the next one the coordinates will also change to the middle point of the new cell.

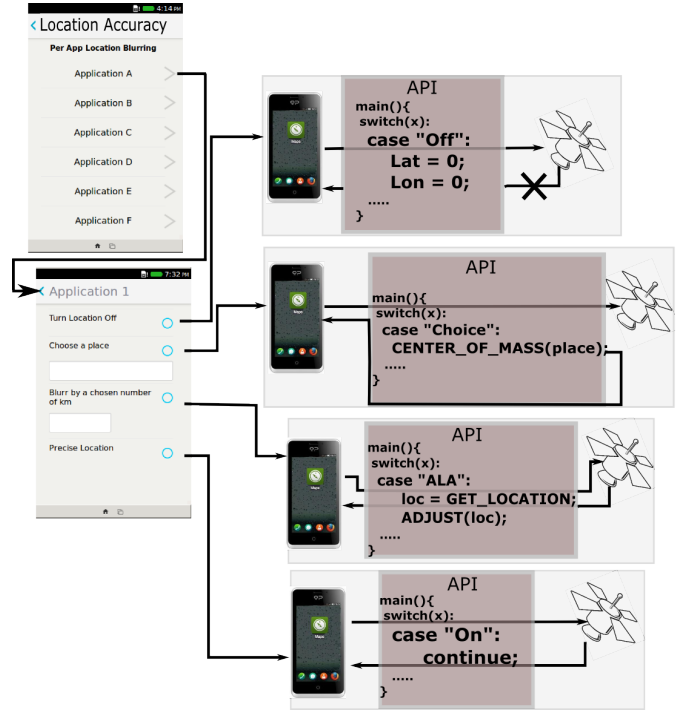

Fig. 4: Simplified mechanism of Adjustable Location Accuracy. We provide the decison on a global and per-app basis. User can choose a separate setting for each app: "no location", when the Latitude and Longitude are set to null, "adjusted" where we use fuzzing algorithm, "custom" where the location is set to a pre defined position, and "precise".

*E. Advanced Secondary-User Mode*

Elements under protection of the Secondary-User Mode can be divided into three groups: apps, data and resources. The applications installed on the phone carry a lot of information about the owner of the device. Suffice it to say, advertisement libraries which allow for targeted advertising send out to the potential clients not only the location, gender, or age of the device's user, but also the list of installed applications [29]. For that reason in the Advanced Secondary-User Mode (SUM) users can decide which apps will be not only accessible, but also visible on the phone. This means that after entering the Secondary-User Mode the apps will disappear from the screen, and from the internal search engine. There is a list of pre-defined apps that will always be removed, like the Settings and the Privacy Dashboard (since the person using our phone should not be allowed to change any options).

Upon entering the Secondary-user mode, the data stored in each of defined elements is substituted with an empty list, just as if no-one ever used the phone. Once the phone is handed back to the owner, the original information is restored. The databases include Contacts, Call History, SMS History, E-mails, Photos, Browser History. Lastly there is a possibility of limiting the access to the resources, like WiFi, Cellular Data etc.

The Secondary-User mode covers multiple use-cases. On one hand it may be used as a "guest mode" to hand over a device for a short period of time. Secondly, it may be used as a permanent solution of parental control – here the parent would create a profile for the child and lock it down. This way the functions can be unlocked as the child grows. Lastly, when properly handled, Secondary-user Mode could be a mobile equivalent of "incognito mode" in web browsers.

## VII. USER STUDY

Having the prototype implemented, that based on our initial research, next step in the user-centric approach was to conduct a structured usability lab study of 26 users. We were mostly interested in answering three research questions:
- Does our Privacy Dashboard bridge the gap, and meets users' expectations?
- What are users' learnability, performance and satisfaction using the privacy features?
- Does the Guided Tour work? Does it actually have impact on people's awareness? Do they accept it?

We started by recording the time required to go through the Guided Tour, as well as looking into the users' attitude and behavior while configuring privacy settings. Next we verified the ease of learning and use of privacy settings, along with the peoples' satisfaction of performing the task, global satisfaction connected to the features and their comments. Each participant learned and performed tasks on an Alcatel One Touch phone with Firefox OS operating system installed. Prior to the study, we have obtained approval in a form of exempt protocol and promised the results would be only used for research.

In total, we had 38% males and 62% females. Their ages were well distributed. We had no participants under 17 or above 55 years old. Most of them (38.5%) graduated from a collage, and 30.8% of them were still in college. 19.2% had a higher education level like a master degree. We had only one participant who has never used a smartphone before.

Based on the background verification, we believe that the opinions expressed and behaviors performed by the participants are representative across age groups, genders, and the feature phone users in smartphone environment.

## VIII. EVALUATION

To eliminate the bias towards difficulties using settings caused by users' own cognition and actions, we asked participants to respond to the questions with regard to using settings in general. We have used the seven-point Likert scale [30] to formulate subjective responses (1- Strongly disagree, 2- Disagree, 3- Somehow disagree, 4- Neither agree nor disagree, 5- Somehow agree, 6- Agree, 7- Strongly agree). Overall, we found that people didn't agree that it is difficult to understand categories and functions in Settings on their smartphone (Median = 2). Moreover they did not have problems with locating specific settings on the device(Median = 2). Interestingly, users did not admit to having difficulties with finding help when stumbling on a problem(Median = 3).

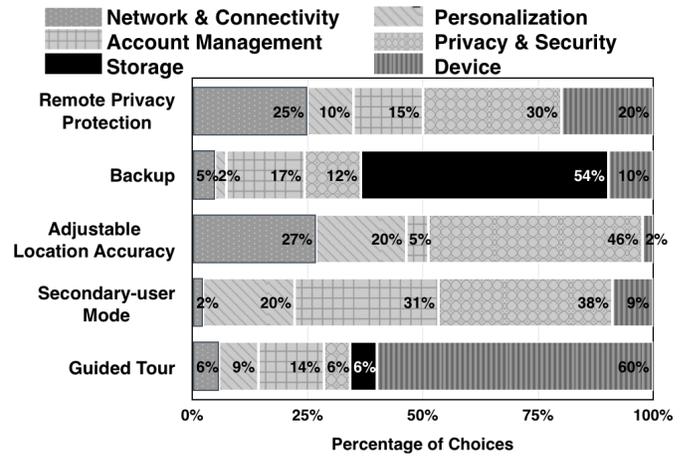

Fig. 5: Categories and Features. The figure presents the way participants categorized each of the features included in the Privacy Dashboard. Not all of them were considered privacy oriented.

### A. Categorization of the features

In the next step we have asked the participants to fill out a pre-test questionnaire. They were supposed to assign features to categories, decide to which privacy level each feature belongs, grade different wireframe options based on good design and bad design, and express comments.

First, we asked the users to put each privacy feature into a different category. We have selected these basing on the elements available in the settings of FirefoxOS: Network & Connectivity, Personalization, Account Management, Privacy & Security, Storage and Device. Each feature could be assigned more than once. We present the results in the Figure 5. The values in the chart represent the percentage of choices instead of participants. We have calculated the data based on the number of choices. The Remote Privacy Protection (RPP), in 30% of times categorized as Privacy & Security, while 25% of choices would connect this feature to Network & Connectivity. When asked to give justification, participants suggested that RPP requires the connection to the network to enable the remote control. 54% of choices were made to assign Backup to the category of Storage. Although next highest result, only 12% of the results put it in the category Privacy & Security. 46%, and 38%, respectively, added both Adjustable Location Accuracy (ALA) and Secondary-user Mode (SUM) into the category of Privacy & Security. ALA was also categorized as Network & Connectivity by 27%, while SUM was placed in the Account Management in 31% of cases. Participants reasoned that the real location information was identified depending on the connectivity of GPS and dispersed via network, thus the N&C choice. People were most consequent when it came to the Guided Tour. 60% choices stated that Guided Tour should belong to Device category.

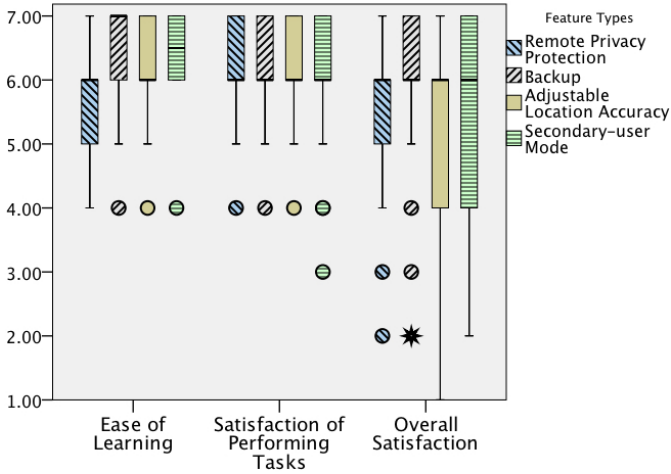

Fig. 6: Learnability, performance and satisfaction of Privacy Features. Dots represent the outliers, while stars denote extreme score.

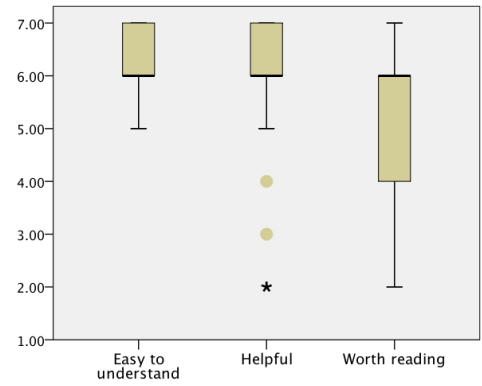

Fig. 7: Ease of use of Guided Tour. Users were asked to state if it was "Easy to understand", "Helpful" and "worth reading". Dots represent outliers while stars denote extreme scores.

*B. General Satisfaction with Privacy Features*

We asked participants to respond to the Likert questionnaire items with regard to the ease of learning, satisfaction of performing tasks and overall satisfaction with the four features (Remote Privacy Protection, Backup, Adjustable Location Accuracy and Secondary-user Mode). We used seven-point scale to formulate responses. Results showed that all participants thought it was easy to learn, they were satisfied with performing the tasks and with features overall. The median scores are all above 6. However, as shown in the Figure 6, more participants gave higher marks towards the ease of learning results, rather than the satisfaction of performing tasks and the overall satisfaction, which reflects on the Adjustable Location Accuracy and the Secondary-user Mode.

We asked users why do they give such scores. From the comments we saw that users found the broad choice and flexibility rather annoying. They did not like the idea of making choice for each app separately. The study clearly showed that the approach of global setting with an exception list, or an opt out mechanism is much more efficient and intuitive. People are used to exploratory study, so they do not find it difficult to discover different options. We have also decided to add a search function to the exception list. Users reported that they clearly know which apps they would like to choose, and a search mechanism speeds up the process.

*C. Guided Tour*

We used the same semantic differential technique to obtain users' feedback on the Guided Tour. The answers were formed by a seven-point scale of semantic differential technique with paring of "Difficult/Easy", "unhelpful/helpful" and "Not worthy/Worthy".

*1) General:* Overall, as shown in the Figure 7, participants thought that Guided Tour was easy to understand (Median = 6), helpful (Median = 6), and worth reading (Median = 6). During the experiment, we asked the users to read the Guided Tour before learning the features. They explored it by themselves and spent the average reading time in the lab as 3m45s (Standard Deviation: 1m45s). Comments were very positive and encouraging: "It was very intuitive", "Pictures help as well", and "Because I am a beginner of this function, I get background information and useful explanations".

*2) Educational aspect of the Guided Tour:* We also wanted to know how successful we were in the task of educating the users on the privacy threats. We have designed and developed eight new statements to evaluate privacy concern of users. These were divided into two, four-statement groups. First part was answered at the stage of background questionnaire, and second was evaluated at the end in the self-report questionnaire:

1) I am afraid what happens to my information in my smartphone if I cannot find it or it is broken.
2) I do not like lending my phone because I do not want anyone to access to information in my smartphone.
3) Providers of apps handle the information they collect about users in an improper and distrust way.
4) Smart phone users have lost control over how data is collected by providers of apps.
5) I found that my personal information in my phone will be not safe if I cannot find my phone or it is broken.
6) I found that my personal information in my phone will be not safe if I lend my phone to someone.
7) I found that my personal information in my phone will be used in a way I feel improper if third-party apps that I installed get it.
8) I found that some of my sensitive personal information in my phone is being obtained/sought by many apps that I installed.

We adopted seven-point Likert scale to formulate the responses, anchored with "strongly disagree" and "strongly agree". These statements do not base on any existing privacy concerns, but are rather created based on the smartphone privacy and activities related to the losing a phone, backup,

| Statements compared | $r_\tau$ | p |
|---|---|---|
| No. 1 with No. 5 | 0,356* | 0,028 |
| No. 2 with No. 6 | 0.476** | 0,003 |
| No. 3 with No. 7 | 0,372* | 0,022 |
| No. 4 with No. 8 | 0,032 | 0,848 |

Fig. 8: Correlation between the answers given before (to statements 1 to 4) and after going through the Guided Tour (statements 5 to 8). Compared were similar statements: statement No. 1 and No. 5, No. 2 and No. 6, No. 3 and No. 7, No. 4 and No. 8. * Correlation is significant at the 0.05 level (2-tailed) ** Correlation is significant at the 0.01 level (2-tailed)

and personal information leakage. We prepared them to be similar, yet not identical. Visual inspections of the histograms showed that the data was not normally distributed. Thus, we have adopted the Kendall's tau non-parametric test [31] to find the correlation between the two groups of answers. Results showed that the first set of answers was positively strongly correlated to the second one, except ones given to statements 4 and 8. As shown in the Figure 8, there was a significant relationship between statement 1 and 5, statement 2 and 6, statement 3 and 7 respectively. This means that the users increased their awareness of privacy problems after taking the Guided Tour.

*3) Changes in Perception of Privacy:* After confirming the positive correlation, we wanted to know the difference of privacy concerns after using Guided Tour and features in Privacy Dashboard. We only used the set that had a significant positive relationship. Due to the lack of normal distribution, we used the Wilcoxon signed ranks test [32] to analyze the data, which is a non-parametric version of a paired t-test on non-normal distributions, working on the data collected using a within-group design. We have found that the privacy concern increased after usage of the tool. However, that change was not as high as we have hoped, thus we have decided to slightly change and extend the Guided Tour.

### D. Discussions and Implications for Design

Our goal was to better understand how users perceive the Privacy Dashboard, and if the features we introduced satisfy their needs. Overall, we found that people thought not all of our features are connected to privacy. They put each feature into different categories, not limited to privacy & security category. Results showed that participants preferred to group five features as Privacy Dashboard and put this panel on the Home Screen, or place them in both the Settings and the app. We believe that this is connected to the ease of access – an app on the Home Screen is easier to find and relate to, than distributed settings.

Second, from the data of users' learnability, performance and satisfaction using privacy features, results showed that all participants thought it was easy to learn, they were satisfied with performing the tasks respectively and with features overall. However we found that over-flexibility is as bad as lack of choice – users were lost in the amount of information in the Adjustable Location Accuracy and Secondary-user Mode cases. For that reason we have changed our design to be more concise.

We are still not fully satisfied with the results of the Guided Tour usage. Although people expressed it is easy to use and they acted that it is worthy reading throughly, the increase of the awareness was not as high as predicted.

*1) The settings beyond the Settings:* Once we start empowering the users, we always need to make sure that we also educate and help them with navigation in the new ecosystem. That is why it was important to us to verify what is the most intuitive placement of the Privacy Dashboard, and how should we place it with respect to the settings. Our study showed that gathering all privacy settings under the umbrella of a single Privacy Dashboard app is the best arrangement.

*2) Educating, advertising and tutoring:* We recommend to provide tutorials to explain the usage of new privacy features like Adjustable Location Accuracy, Secondary-user Mode and use of commands in Remote Privacy Protection. The current Guided Tour did not provide as much help as we hoped for, thus we decided to extend the feature with three main focus areas. Instead of just educating the user the Guided Tour should also advertise and tutor. Especially beginners need more explanation on how to use the new elements and to be guided to notice them.

*3) Personalization of settings:* Finally, to let users adapt themselves to app ecosystem and personalization of settings, we consider improving the functions for Adjustable Location Accuracy and Secondary-user Mode, in the way of supporting more sorting methods like mostly used, and recommendation engine, based on user's own data. Concerning Secondary-user Mode, to go a step further, we consider to provide the fast access function which can assist creating new profiles and reusing these profiles for users.

### IX. CONCLUSIONS

In this paper we presented the Privacy Dashboard: a tool that we have created using method that we call "User-centric privacy development". We begun our work by doing a wide scale study to verify and judge what is the current state of privacy awareness of an average user. As an outcome of this study we had an overview of what do people expect, know and hope for. Next, having defined the gap between users' expectations and the capabilities that devices have, we have designed and developed a prototype solution to meet the requirements. Having the initial solution we have conducted another study, to analyze how the prototype fits into the users' needs. Based on that we have created a second iteration of improvements and a final version of the tool. The feedback from the users showed that the adaptation rate of features developed is very high. It has also lead us to the changes in the presentation of Location Accuracy and Secondary-user Mode as well as the layout of the Guided Tour. This way we have conducted research much wider, than would normally be done when developing a solution, and made sure that our work is not only of high academic quality, but also can be introduced on product-scale and will be easily adapted. We believe that involving users in the research and design phase, and building solutions that meet their needs is crucial to bring the academic work closer to the people.

As a result we have created a tool that addresses five basic concerns of users – disclosure of their location, sharing data with people who temporary use their phone, requirement of registering in a central data-base in order to have a "Find My Device" feature, lack of choice when it comes to backup methods. In addition to the improvements to users' privacy that our tool offers, and the fact it fits into their expectations, we have also gathered in-depth understanding of peoples' behavior patterns in the field. We have presented those results in the paper. Our prototype runs on Firefox OS, with a plan to investigate and develop each feature, hoping that we will be able to collaborate towards introducing it on a wide scale.

## REFERENCES

[1] M. Stanley, "The mobile internet report," Morgan Stanley, Tech. Rep., 2014.
[2] B. Gellman and L. Poitras, "U.s., british intelligence mining data from nine u.s. internet companies in broad secret program," *The Washington Post*, June 7, 2013.
[3] J. Appelbaum. (2013, 2014) Nsa-programm prism. [Online]. Available: http://www.spiegel.de/international/search/index.html?suchbegriff=appelbaum
[4] Multiple. Articles on edward snowden. [Online]. Available: http://www.theguardian.com/us-news/edward-snowden
[5] N. C. Shaw, W. H. DeLone, and F. Niederman, "Sources of dissatisfaction in end-user support: An empirical study," *SIGMIS Database*, vol. 33, no. 2, pp. 41–56, Jun. 2002. [Online]. Available: http://doi.acm.org/10.1145/513264.513272
[6] T. Lowdermilk, *User-Centered Design: A Developer's Guide to Building User-Friendly Applications*. O'Reilly Media Inc., 2013.
[7] A. Hang, E. von Zezschwitz, A. De Luca, and H. Hussmann, "Too much information!: User attitudes towards smartphone sharing," in *Proceedings of the 7th Nordic Conference on Human-Computer Interaction: Making Sense Through Design*, ser. NordiCHI '12. New York, NY, USA: ACM, 2012, pp. 284–287. [Online]. Available: http://doi.acm.org/10.1145/2399016.2399061
[8] J. C. J. Keng, T. K. Wee, L. Jiang, and R. K. Balan, "The case for mobile forensics of private data leaks: Towards large-scale user-oriented privacy protection," in *Proceedings of the 4th Asia-Pacific Workshop on Systems*, ser. APSys '13. New York, NY, USA: ACM, 2013, pp. 6:1–6:7. [Online]. Available: http://doi.acm.org/10.1145/2500727.2500733
[9] L. Jedrzejczyk, B. A. Price, A. K. Bandara, and B. Nuseibeh, "On the impact of real-time feedback on users' behaviour in mobile location-sharing applications," in *Proceedings of the Sixth Symposium on Usable Privacy and Security*, ser. SOUPS '10. New York, NY, USA: ACM, 2010, pp. 14:1–14:12. [Online]. Available: http://doi.acm.org/10.1145/1837110.1837129
[10] B. Wentz and J. Lazar, "Are separate interfaces inherently unequal?: An evaluation with blind users of the usability of two interfaces for a social networking platform," in *Proceedings of the 2011 iConference*, ser. iConference '11. New York, NY, USA: ACM, 2011, pp. 91–97. [Online]. Available: http://doi.acm.org/10.1145/1940761.1940774
[11] I. Shklovski, S. D. Mainwaring, H. H. Skúladóttir, and H. Borgthorsson, "Leakiness and creepiness in app space: Perceptions of privacy and mobile app use," in *Proceedings of the 32Nd Annual ACM Conference on Human Factors in Computing Systems*, ser. CHI '14. New York, NY, USA: ACM, 2014, pp. 2347–2356. [Online]. Available: http://doi.acm.org/10.1145/2556288.2557421
[12] J. Lin, *Understanding and Capturing People's Mobile App Privacy Preferences*. Pittsburgh, PA, USA: Carnegie Mellon University, 2013, aAI3577905.
[13] M. J. Keith, S. C. Thompson, J. Hale, P. B. Lowry, and C. Greer, "Information disclosure on mobile devices: Re-examining privacy calculus with actual user behavior," *Int. J. Hum.-Comput. Stud.*, vol. 71, no. 12, pp. 1163–1173, Dec. 2013. [Online]. Available: http://dx.doi.org/10.1016/j.ijhcs.2013.08.016
[14] D. Biswas, I. Aad, and G. P. Perrucci, "Privacy panel: Usable and quantifiable mobile privacy," in *Proceedings of the 2013 International Conference on Availability, Reliability and Security*, ser. ARES '13. Washington, DC, USA: IEEE Computer Society, 2013, pp. 218–223. [Online]. Available: http://dx.doi.org/10.1109/ARES.2013.29
[15] M. Pirker and D. Slamanig, "A framework for privacy-preserving mobile payment on security enhanced arm trustzone platforms," in *Proceedings of the 2012 IEEE 11th International Conference on Trust, Security and Privacy in Computing and Communications*, ser. TRUSTCOM '12. Washington, DC, USA: IEEE Computer Society, 2012, pp. 1155–1160. [Online]. Available: http://dx.doi.org/10.1109/TrustCom.2012.28
[16] A. Pingley, W. Yu, N. Zhang, X. Fu, and W. Zhao, "A context-aware scheme for privacy-preserving location-based services," *Comput. Netw.*, vol. 56, no. 11, pp. 2551–2568, Jul. 2012. [Online]. Available: http://dx.doi.org/10.1016/j.comnet.2012.03.022
[17] J. Kong, "Anonymous and untraceable communications in mobile wireless networks," Ph.D. dissertation, 2004, aAI3147729.
[18] Y. Guo, L. Zhang, and X. Chen, "Collaborative privacy management: Mobile privacy beyond your own devices," in *Proceedings of the ACM MobiCom Workshop on Security and Privacy in Mobile Environments*, ser. SPME '14. New York, NY, USA: ACM, 2014, pp. 25–30. [Online]. Available: http://doi.acm.org/10.1145/2646584.2646590
[19] Apple. Our commitment to customer privacy doesn't stop because of a government information request. [Online]. Available: https://www.apple.com/privacy/government-information-requests/
[20] E. Kain. The nsa reportedly has total access to the apple iphone. [Online]. Available: http://www.forbes.com/sites/erikkain/2013/12/30/the-nsa-reportedly-has-total-access-to-your-iphone/
[21] J. Zdziarski. Identifying back doors, attack points, and surveillance mechanisms in ios devices. [Online]. Available: http://www.zdziarski.com/blog/wp-content/uploads/2014/08/Zdziarski-iOS-DI-2014.pdf
[22] S. Rosenblatt, "Why android won't be getting app ops anytime soon," *CNET*, December 19, 2013.
[23] E. Parliament, "Directive 2002/58/ec," *Official Journal L 201 , 31/07/2002 P. 0037 - 0047*, 2002.
[24] S. Consolvo, I. E. Smith, T. Matthews, A. LaMarca, J. Tabert, and P. Powledge, "Location disclosure to social relations: Why, when, and what people want to share," *CHI'05*, 2005.
[25] Apple. ios: About guided access. [Online]. Available: support.apple.com/en-us/HT202612
[26] K. Parrish. Google: "guest mode" for android, more nexus devices soon. [Online]. Available: http://www.tomshardware.com/news/Sundar-Pichai-Guest-Mode-Chrome-D11-HTC-one,22853.html
[27] B. A. Mellers, A. Schwartz, and A. D. J. Cooke, "Judgment and decision making," *Annual Review of Psychology*, no. A. Schwartz 2 , and A. D. J. Cooke.
[28] R. Kumar. It's time for youth and governments to fall in love. [Online]. Available: http://blogs.worldbank.org
[29] D. Malandrino, A. Petta, V. Scarano, L. Serra, R. Spinelli, and B. Krishnamurthy, "Privacy awareness about information leakage: Who knows what about me?" in *Proceedings of the 12th ACM Workshop on Workshop on Privacy in the Electronic Society*, ser. WPES '13. New York, NY, USA: ACM, 2013, pp. 279–284. [Online]. Available: http://doi.acm.org/10.1145/2517840.2517868
[30] I. Allen and C. A. Seaman, "Likert scales and data analyses,," *Quality Progress, vol. 40, no. 7*, 2007.
[31] A. Field, "Discovering statistics using ibm spss statistics," *Sage*, 2013.
[32] R. Ott and M. Longnecker, *An Introduction to Statistical Methods and Data Analysis*. Cengage Learning, 2008.